\documentstyle[12pt,psfig]{l-aa}

\def\degmark{^\circ}
\def \rsun {\ifmmode$R$_{\odot}\else R$_{\odot}$\fi}

\def \hcm {\hbox {\ifmmode $ H atoms cm$^{-2}\else H atoms cm$^{-2}$\fi}}
\def \src {MCG-6-30-15}
\def\approxgt{\mathrel{\hbox{\rlap{\lower.55ex \hbox {$\sim$}}
        \kern-.3em \raise.4ex \hbox{$>$}}}}
\def\approxlt{\mathrel{\hbox{\rlap{\lower.55ex \hbox {$\sim$}}
        \kern-.3em \raise.4ex \hbox{$<$}}}}
\newcommand {\ginga} {{Ginga}}
\newcommand {\rosat} {{ROSAT}}

\newcommand {\sax} {{BeppoSAX}}

\newcommand {\exosat} {{EXOSAT}}
\newcommand {\asca} {{ASCA}}

\newcommand {\chisq} {$\chi ^{2}$}

\psfigurepath{/usr4/users/aorr/mcg-6-30-15}

\begin{document}

\thesaurus{ (11.09.1; 11.19.1; 13.25.2)}

\title{Soft X-ray observations of the complex warm absorber in \src~with \sax}

\author{Astrid Orr\inst{1} \and S. Molendi\inst{2} \and F. Fiore\inst{3} 
\and P. Grandi\inst{4} \and A.N. Parmar\inst{1} \and Alan Owens\inst{1}}

\institute{Astrophysics Division, Space Science Department of ESA, 
ESTEC, P.O. Box 299, 2200 AG Noordwijk, The Netherlands
\and
Instituto di Fisica Cosmica e Tecnologie Relative C.N.R., Via Bassini 15, I-20133 
Milano, Italy 
\and
\sax\ Science Data Center, c/o Nuova Telespazio,
Via Corcolle 19, I-00131 Roma, Italy 
\and
Instituto di Astrofisica Spaziale C.N.R., Via Enrico Fermi 21-23, I-00044 
Frascati, Italy
}

\date{Submitted to Astronomy and Astrophysics}
\offprints{A. Orr: aorr@astro.estec.esa.nl}
\maketitle

\begin{abstract}
We report on soft X-ray observations of 
the Seyfert~1 galaxy \src\ with the Low  and Medium Energy Concentrator 
Spectrometers on board \sax.
The time averaged 0.1--4~keV spectrum shows evidence for 
a complex warm absorber. K-edges of highly 
ionized oxygen (O {\sc vii} and O {\sc viii}) can only partly describe the soft X-ray
spectrum below 2 keV. A spectral variability study reveales 
large and rapid changes in absorption at the energy of the O {\sc viii} K-edge
and in the continuum shape and count rate. 
These fluctuations are not simply correlated.
\end{abstract}

\keywords{Galaxies: individual: (\src) $-$ Galaxies: Seyfert $-$
X-rays: galaxies}

\section{Introduction}
\label{sec:intro}

The well studied and nearby (z = 0.008) Seyfert 1 galaxy \src~displays 
complex X-ray emission. 
Strong and rapid variability, as well as intrinsic absorption and possibly
a ``soft excess'' component were first detected with \exosat~ 
(Pounds et al. 1986). \src~was among the first objects in which 
a spectral ``hump'' was observed above 10 keV (Pounds et al. 1990) 
suggesting the reprocessing of hard X-rays by cold, optically thick reflecting 
matter.
The \rosat~Position Sensitive Proportional Counter
(PSPC; Nandra \& Pounds 1992) provided evidence for the presence of
highly ionized ``warm'' absorbing medium close to the X-ray continuum source. 
This absorption was detected as a feature at 0.8 keV,
consistent with a blend of O {\sc vii} and O {\sc viii} K-edges. 
\asca~ performed further observations of \src~ (Fabian et al. 1994;
Reynolds et al. 1995; Otani et al. 1996), which indicate
short term variability of the warm absorber. Reynolds et al. (1995) and 
Otani et al. (1996) report
dissimilar variability patterns in the depth of the O {\sc vii} and O {\sc viii} 
K-edges and conclude that
these features must originate at different distances from the central engine.

\section{Observations}
\label{sec:obs}

\src\ was observed by the \sax~satellite as part of the 
Science Verification Phase
between 1996 July 29, 15:56 and August 03, 03:15~UT. 
The data reported here were obtained using the Low and Medium Energy 
Concentrator Spectrometers (LECS \& MECS). The LECS and MECS are position 
sensitive gas scintillation proportional counters, sensitive in the
energy range 0.1--10 keV and 1.3--10 keV respectively  with  circular fields 
of view of diameters 37' and 56' (LECS: Parmar et al. 1997; MECS: Boella et al. 
1997).
The LECS has good spectral resolution at 
energies $\approxlt 0.5$~keV, where instruments such as the Solid State Imaging
Spectrometer (SIS) on 
\asca~are not sensitive (see Tanaka et al. 1994) 
and where instruments such as the 
\rosat\ PSPC (0.1--2.5~keV; 
Tr\"umper 1983) have only moderate spectral resolution.
Above 1.8 keV~the effective area of the MECS is larger than that of the LECS. 
The total on-source exposure is 49.4~ksec for the LECS and 184~ksec for the 
MECS. 
Data from the LECS and the MECS were prepared in a similar way
(the MECS data are discussed in further detail by Molendi et al. 1997).
\begin{figure}[htb]
\hbox{\psfig{figure=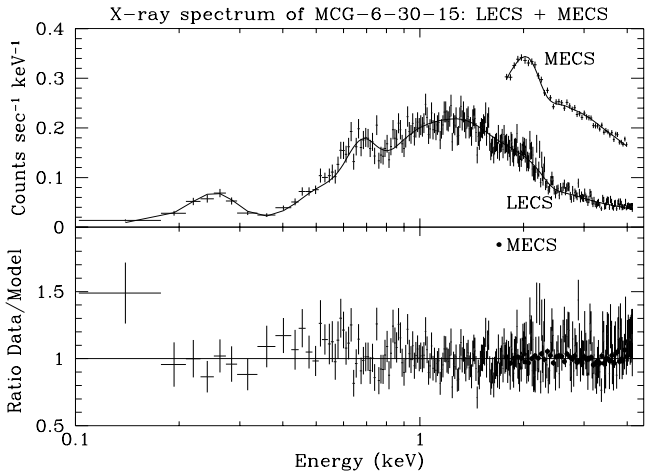,height=6.3cm,width=9cm,angle=-90} }
\caption{\protect \small The combined LECS/MECS X-ray spectrum of MCG-6-30-15. 
The 1.8--4 keV data set is 
from the MECS. Top panel, observed mean 0.1--4 keV spectrum
fitted by a power-law, an absorption edge and galactic plus excess neutral 
absorption (model 2 of Table \ref{averfit}).
Lower panel, ratio of data and model. The residuals
are discussed in the text}
 \label{spec}
\end{figure}
LECS data were processed using the SAXLEDAS 1.4.0 data analysis package 
(Lammers 1997).
The mean spectrum was extracted within a radius of 4$'$ of the source
centroid, corresponding to 55 \% of the encircled energy at 0.28~keV 
in the LECS. 
Light curves and the spectra for several shorter time intervals were 
obtained using an 8$'$ LECS extraction radius which includes 95\% of the 
0.28~keV photons. 
Background subtraction was
performed using a 46~ksec exposure obtained by summing 
blank field observations at galactic latitudes close to that of 
\src~($\Delta b < 5.3 \degmark$). 
After background subtraction and with a 4 $'$ extraction radius, the LECS 
count rate is $0.5243 \pm 0.0033$~s$^{-1}$ and that of the MECS is
$1.0370 \pm 0.0025$~s$^{-1}$.

\section {Spectral fits}
\label{spectrum}

Fits to LECS and MECS data were performed 
in order to describe the mean spectrum of \src, 
as well as to investigate possible spectral variability between selected 
time intervals. Fe K$\alpha$ emission contributes significantly to the 
flux between 4--10~keV (Tanaka et al. 1995; Iwasawa et al. 1996; 
Molendi et al. 1997). Therefore 
data above 4~keV are not considered for the present study which focuses on 
the soft X-ray range, but are discussed by Molendi et al. (1997).
The simultaneous fitting of LECS and MECS observations imposes
strong constraints on the spectral energy distribution in the common
energy range from 1.8--4.0 keV while benefiting from the good spectral
resolution of the LECS at lower energies. All fit models include a 
constant multiplicative factor to account for the LECS/MECS inter-calibration. 
This factor varied no more than 2\% between the different models fitted.
 
A minimum of at least 20 source counts per spectral bin
was chosen to ensure meaningful $\chi^2$ values. 
Uncertainties correspond to $\Delta \chi ^2 + 2.706$, unless otherwise stated. 
\begin{figure}[htb]
\hbox{\psfig{figure=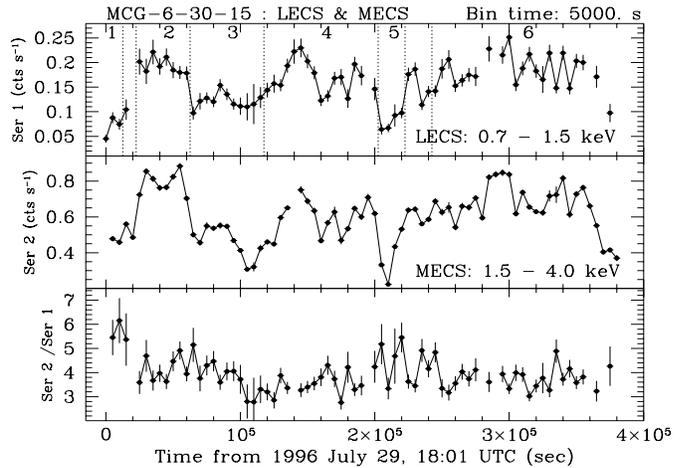,height=6.3cm,width=9cm,angle=-90} }
\caption{\protect \small 
Top: 0.7--1.5 keV LECS light curve of MCG-6-30-15 
within a radius of 8$'$ of the source centroid. 
The epochs chosen for the study of spectral 
variability are numbered from 1 to 6.
Middle: 1.5--4.0 keV MECS light curve (4$'$ radius).
Bottom: ratio of 1.5--4.0 keV counts over
0.7--1.5 keV counts.
The corresponding background count rates are
negligible; LECS: ($5.40 \, \pm 0.38) \, 10^{-3}$ s$^{-1}$;
            MECS: ($1.45 \, \pm 0.13) \, 10^{-3}$ s$^{-1}$
}
 \label{light}
\end{figure}

\subsection{The mean spectrum}
\label{subs:mean_spec}

\begin{table*}[ht] 
\begin{center}
\caption{\protect \small Results of combined LECS/MECS fits to the mean 
0.1 to 4 keV spectrum of \src.
All fits include photo-electric 
absorption with $N_{Gal} = 4.06 \times 10^{20}$ cm$^{-2}$. 
The uncertainties are $\Delta \chi ^2 + 2.706$.
Rest frame energies are given for the quoted absorption edges and emission 
lines.
Parameters without uncertainty values were fixed during the fitting}
\label{averfit}
\begin{tabular}{llllllll} \hline
\multicolumn{2}{l}{model} & excess $N_H$ & photon index & E$_{edge}$ & ${\tau}_{edge}$ & $E_{line}$
 & ${\chi}^2 $/d.o.f \\
  &    & ($10^{20}$ cm$^{-2}$)  &  $\Gamma $  & (keV)  & & (keV)  &  \\ \hline 
1 & PL  & ... & 1.68 $\pm~0.01$  & ... & ... & ... & 707.1/400 \\
& & & & & & & \\
2 & PL+edge & 1.86 $\pm~{0.30}$ & 1.89 $\pm~{0.02}$ & 0.77 $\pm~{0.02}$ & 0.81 $\pm~{0.10}$ & ... & 418.4/396 \\
&  & & & & & & \\
3 & PL+2 edges & 1.94 $\pm~{0.31}$ & 1.90  $\pm~{0.02}$ & 0.74 & 0.65 $\pm~{0.14}$ & ... & 415.9/396 \\
&  &  &  & 0.87 & 0.22 $\pm~{0.11}$  & & \\  
&  & & & & & & \\
4 & PL+2 edges & 1.89 $\pm~{0.33}$ & 1.88 $\pm~{0.03}$ & 0.74 
& 0.66 $\pm~{0.14}$ & 0.59 $\pm~{0.03}$  & 403.8/394 \\ 
& + line & & & 0.87 & 0.14  $\pm~{0.12}$ & & \\  
&  & & & & & & \\
5 & PL+3 edges & 2.36 $\pm~{0.38}$ & 1.95 $\pm~{0.03}$ & 0.74 
& 0.91 $\pm~^{0.18}_{0.14}$ & ...  & 398.1/394 \\ 
&  & & & 0.87 & 0.03 $\pm~{0.15}$ &  & \\
&  & & & 1.12 $\pm~{0.10}$ & 0.19 $\pm~{0.08}$ &  & \\  \hline
\end{tabular}
\end{center}
\end{table*}

Table \ref{averfit} describes the various fits to the mean 0.1 to 4 keV 
spectrum of \src. 
A simple power-law fit 
with photo-electric absorption fixed at the galactic hydrogen
column density $N_{Gal}= 4.06 \times 10^{20}$ cm$^{-2}$ (Elvis et al. 1989) 
gives a poor fit (\chisq~= 707.1 with 400 degrees of freedom, dof) 
with large residuals below 2 keV.

Including excess photo-electric absorption and an absorption edge component
to account for a blend of O {\sc vii} and O {\sc viii} gives a significantly better fit
(\chisq~=418.4 with 396 dof; Fig. \ref{spec}).
Allowing for excess absorption improves 
the quality of the fit in all models considered. 
This was also the case for \asca~observations (Otani et al. 1996).

We have tried to fit the warm absorber 
in the LECS mean spectrum with two absorption edges.
However the fits do not require two separate edges around the 
physical rest frame energies of the O {\sc vii} and O {\sc viii} K-edges: 
0.74 and 0.87 keV.
This is most likely due to a combination of effects, such as
the temporal variability of the optical depths of the edges (reported by
Fabian et al. 1994, Reynolds et al. 1995 and Otani et al. 1996)
which can bias the energy of the O {\sc vii} and O {\sc viii} edge blend.
Also, the energy resolution of the LECS in this energy range is poorer 
than that of the SIS on \asca.
Little improvement in fit quality is obtained with two edges fixed at these 
energies (\chisq~= 415.9 with 396 dof).

The fit residuals for all models mentioned above show complex structure
below 1 keV. Otani et al. (1996) included an {\it ad hoc} Gaussian emission 
line at 0.6 keV to improve their fits to the mean \asca~ spectrum. 
A Gaussian line also provides a better fit to the 
LECS mean spectrum (E$_{rest} = 0.59\; \pm \; 0.03$
keV; EW = 44.5 $\pm$ 30.7 eV; \chisq~= 403.8 with 394 dof). 
The energy of the line is consistent
with a blend of O {\sc vii} and O {\sc viii} recombination lines at E$_{rest}$ = 0.57 and
0.65 keV, respectively, which are unresolved by the LECS. The resulting
reduction of \chisq~ is significant at the 99\% level on the basis of an
F-test. 
The line is narrower than the resolution of the LECS (which is
140 eV at 0.6 keV) and when
left as a free fit parameter the line width, $\sigma$, becomes zero. 
Consequently the line width has been fixed at 10 eV. 

The parameters of the warm absorber component and in particular the optical 
depths depend on the shape of the underlying continuum. As can be seen in 
Table \ref{averfit} the power-law index remains almost constant for models
2 to 4, since it essentially constrained by the data above 1~ keV. 
However, this spectral range may not be representative
of the underlying continuum flux. In particular a blend of absorption edges
may considerably affect the spectral range up to 3 keV (Ferland 1996; 
Nicastro et al. in preparation), 
thereby causing the local spectrum to be harder than the underlying
continuum. 
In order to test this hypothesis we have added a Ne {\sc ix} absorption edge 
 to the fit (model no. 5 in Table \ref{averfit}). 
Its energy is E$_{rest}$ = 1.12 $\pm \; 0.10$ keV which is compatible with the
physical value of 1.20 keV. The decrease in \chisq~
with respect to model no. 3 including only oxygen edges is significant at the 
99\% level on the basis of an F-test. Furthermore the photon index in  
model 5 is marginally steeper than with models 2 -- 4, as expected.

In conclusion, none of the models described above gives an entirely  
satisfactory fit to the mean 0.1--4 keV spectrum of \src, although the 
addition of various components discussed can significantly reduce \chisq. 
In particular, residual flux remains below 0.2 keV. Features below the 
C-edge (0.28 keV) are very sensitive to the subtracted background which 
has a soft component in this spectral range (Hasinger et al. 1993). 
Furthermore the diffuse soft X-ray background map obtained with the 
\rosat~PSPC (Snowden et al. 1995) shows considerable structure. For all these
reasons background subtraction is complex. We have deliberately chosen 
a small extraction radius for the LECS spectra in order to reduce the
contributions from the background.

\subsection{Spectral variability}
\label{subs:var_spec}

Because of the difficulty in obtaining satisfactory fits to the mean spectrum, 
we have investigated whether this is due to spectral variability. The LECS 
0.7--1.5 and MECS 1.5--4.0 keV light curves of \src, as well as their ratio, 
are shown in Fig. \ref{light}. As seen in the previous section, warm
absorption strongly affects the first spectral range, but much less the  
second. 
Flux variations of the order of a factor 4  occured during our 
observation.
In fact, at the start of the observation the 0.7--1.5 keV
count rate increased four-fold in $<$7 hours while the 0.7--4 keV spectrum
was in a ``hard'' state. Such behaviour suggests a change in the warm absorber.
We performed time resolved spectral analysis in order to check which
ion species (${\rm O} {\sc vii}$ or ${\rm O} {\sc viii}$) had varied.
For ease of analysis we selected six ``low'' and ``high'' states, with 
average LECS 0.1--10 keV 
count rates respectively below or above a value of 0.5 cts s$^{-1}$. 
This choice
is by definition inhomogeneous, because of complex time structure
within the time intervals. In particular, some data between intervals 
have been excluded.

\begin{figure}[htb]
\hbox{\psfig{figure=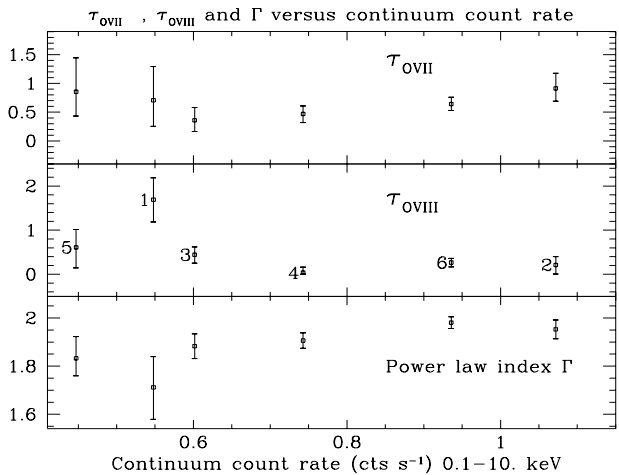,height=6.3cm,width=9cm,angle=-90} }
\caption{\protect \small Photon index, $\Gamma$, $\tau_{{\rm O} 
{\sc vii}}$ and $\tau_{{\rm O} {\sc viii}}$ versus un-absorbed model 
continuum count rate from 0.1 to 10 keV.
The 0.1--4 keV fits use model 3 of Table \ref{averfit}.
Uncertainties are $\Delta \chi ^2 + 1.0 $.
In the middle panel, each point is identified by its epoch number as 
defined in Fig. \ref{light}}
 \label{taur}
\end{figure}

Model 3 is used to fit the time selected LECS+MECS spectra from 0.1--4 keV.
Figure \ref{taur} shows that the optical depths
$\tau_{{\rm O} {\sc vii}}$ and $\tau_{{\rm O} {\sc viii}}$ have different variability patterns.
The depth of the O {\sc vii} edge is marginally consistent with being constant 
throughout the LECS observation, whereas $\tau_{{\rm O} {\sc viii}}$ exhibits significant
variability. 
Its large value during interval 1 ($1.7\pm 0.5$, $1\sigma$ uncertainty) is 
inconsistent with the values at all other epochs, and is the cause of the 
spectral hardening at epoch 1 (Fig. \ref{light}, lower panel). 
Epoch 1 stands out as having a more extreme value of $\tau_{{\rm O} 
{\sc viii}}$
and count rate variability than observed by \asca~ in 1993 and 1994 
(Fabian et al. 1994; Reynolds et al. 1995; Otani et al. 1996).
This epoch is characterized by a very rapid increase in count rate. 
Since the source count rate prior to our observation is unknown, 
the ionizing flux could have been much lower earlier,
leading to the high value of $\tau_{{\rm O} {\sc viii}}$ observed.

\section {Discussion}

Our results indicate that the spectral variability of \src~from 0.1--4 keV 
is due to changes in the
continuum emission as well as modifications of the warm absorber.
These variations have similar timescales, but are not always associated 
with one another.  
For instance during epoch 5 (Figs. \ref{light} and \ref{taur})
the 1.5--4 keV flux undergoes a rapid ($\tau_{var} < 2 \times 10^4$ s) decrease
followed by an equally fast increase. With such rapid changes it is possible
that the absorber is not in photo-ionization equilibrium. 
Indeed, the spectrum of epoch 5 shows no evidence of change in the warm 
absorber. 

Likewise, the spectral softening before high state 6
is not accompanied by any significant change in 
$\tau_{{\rm O} {\sc vii}}$ or $\tau_{{\rm O} {\sc viii}}$.
Spectral softening with increasing flux was also observed in \src~ by
\ginga~in the 2--18~keV band and variously interpreted by several groups
(Matsuoka et al. 1990; Nandra et al. 1990; Fiore et al. 1992).
A more detailed analysis of these variations in the \sax~data is presented 
by Molendi et al. (1997).

There are clear indications that more than one absorption 
edge is present below 1 keV. Models including both an
O {\sc vii} and an O {\sc viii} absorption edge bring little improvement 
to the quality of 
fit over single edge models. However, single edge models are physically
unacceptable because the spectral fits to the time intervals then require 
temporal 
variations  of the edge energy. 
The LECS data show that the optical depth
$\tau_{{\rm O} {\sc vii}}$ did not change significantly during the exposure, 
whereas $\tau_{{\rm O} {\sc viii}}$
varied on a time scale $\leq$ 15 000 seconds (epochs 1 \& 2 in Figs. 
\ref{light} and \ref{taur}). Differences 
between $\tau_{{\rm O} {\sc vii}}$ and $\tau_{{\rm O} {\sc viii}}$ have been 
reported by Reynolds et al. (1995) and Otani et al. (1996). 
However, unlike what was measured with \asca, our data show no simple 
anti-correlation between $\tau_{{\rm O} {\sc viii}}$ and the continuum 
luminosity.
A warm absorber consisting of spatially distinct photo-ionized regions, 
such as the two-zone model applied to \asca~data
(Reynolds et al. 1995; Otani et al. 1996), does not provide an entirely
satisfactory explanation to such behaviour.
A plausible interpretation is that the warm absorber is not in
simple photo-ionization equilibrium with the incident X-ray continuum. 
More complex processes, such as non-equilibrium -- or collisional
photo-ionization, may instead contribute to the soft X-ray spectrum of \src.

\begin{acknowledgements}
The authors wish to thank the \sax~science team members. The \sax~ satellite 
is a joint Italian and Dutch programme. Astrid Orr acknowledges an ESA 
Research Fellowship.

\end{acknowledgements}

\end{document}